# The bumpy road towards iPhone 5c NAND mirroring

Sergei Skorobogatov
University of Cambridge
Computer Laboratory
Cambridge, UK
e-mail: sps32@cam.ac.uk

*Abstract*—This paper is a short summary of a real world mirroring attack on the Apple iPhone 5c passcode retry counter under iOS 9. This was achieved by desoldering the NAND Flash chip of a sample phone in order to physically access its connection to the SoC and partially reverse engineering its proprietary bus protocol. The process does not require any expensive and sophisticated equipment. All needed parts are low cost and were obtained from local electronics distributors. By using the described and successful hardware mirroring process it was possible to bypass the limit on passcode retry attempts. This is the first public demonstration of the working prototype and the real hardware mirroring process for iPhone 5c. Although the process can be improved, it is still a successful proof-of-concept project. Knowledge of the possibility of mirroring will definitely help in designing systems with better protection. Also some reliability issues related to the NAND memory allocation in iPhone 5c are revealed. Some future research directions are outlined in this paper and several possible countermeasures are suggested. We show that claims that iPhone 5c NAND mirroring was infeasible were ill-advised.

*Keywords: Apple iPhone 5c; NAND Flash memory; mirroring attack; hardware security*

## I. INTRODUCTION

Mobile phones, and in particular smart phones, can contain a large amount of personal information: contact history, text messages, location history, access-credentials to online services, financial details, etc. It is therefore hardly surprising that the forensic examination of mobile-device storage has become a significant line of enquiry in many police investigations, and forces around the world operate large laboratories to routinely retrieve and analyze data from the phones of both suspects and victims. At the same time, smartphones are evolving into personal security devices used for financial transactions, with associated user expectations about their physical security. Mobile phone vendors, most notably Apple Inc., have responded by encrypting data stored in non-volatile memory, in order to protect personal data and access credentials against unauthorized recovery of from lost or stolen devices.

Data mirroring is widely used in computer storage when higher reliability of data storage is required. This is a process of copying data from one location to a storage device in real time. As a result the information stored from the original location is always an exact copy of the original data. Data mirroring is useful in recovery of critical data after a disaster. In computer systems mirroring can be implemented as a part of standard RAID (redundant array of independent disks) levels [1]. From the hardware security prospective the process of mirroring could pose a threat as it creates a backup copy of the data that might allow restoring the previous state of the system, for example, with a higher value of password retry counter.

The Apple iPhone 5c went under the spotlight soon after FBI recovered one from a terrorist suspect in December 2015 [2]. In February 2016 the FBI announced that it was unable to unlock the recovered phone due to its advanced security features, including encryption of user data [3]. The FBI first asked the NSA to break into the phone, but they were unable to [4]. As a result, the FBI asked Apple Inc. to create a new version of the phone's iOS operating system that could be installed and run in the phone's random access memory to disable certain security features. Apple refers to this as "GovtOS". Apple declined due to its policy to never undermine the security features of its products. The FBI responded by successfully applying to a United States magistrate judge, Sherri Pym, to issue a court order, mandating Apple to create and provide the requested software [5]. Less than 24 hours before a highly anticipated hearing over access to the phone was set to begin, Justice Department lawyers requested a delay [6]. Later in March the Justice Department has abandoned its bid to force Apple to help it unlock the iPhone saying that they had "now successfully accessed the data" stored on the iPhone in question [7].

At a press conference on 24 March 2016 FBI Director James Comey told reporters that "NAND mirroring" will not be used to get into the terrorist's iPhone 5c, saying "It doesn't work" [8,9].

NAND mirroring was suggested by several technology experts as the most likely way to gain unlimited passcode attempts in iPhone 5c. iPhone forensics expert Jonathan Zdziarski has demonstrated a software-based proof-of-concept of mirroring attack using jailbroken iPhone 5c. Although he did it with a jailbreak, he noted that "no jailbreak is needed to do this", "as the FBI would be physically removing the NAND to copy this data". The FBI Director Comey was not pleased about the piece by saying: "You are simply wrong to assert that the FBI and the Justice Department lied about our ability to access the San Bernardino killer's phone" [10].

So far no one has demonstrated a fully working hardware-based NAND mirroring attack on iPhone 5c. Therefore, this paper is aimed at demonstrating the feasibility of such a process. Although it does not require expensive equipment there were several unexpected traps, pitfalls and obstacles on the way to full success. Most of these challenges are described in this paper.

This paper is organized as follows. Section 2 gives a brief introduction into iPhone 5c hardware and NAND

memory. Section 3 introduces the preparation part, while Section 4 sets out the results. Section 5 discusses limitations and Section 6 possible improvements. The impact of the research is discussed in the concluding section.

## II. BACKGROUND

The underlying security of iOS based devices including iPhone 5c is described in the Apple iOS Security guide [11]. Some forensics experts describe the differences between different versions of devices, for example, the encryption key management in iOS is presented in Figure 1 [12]. The user's passcode together with each device's unique UID key are used to calculate the Passcode key which unlocks the "System Keybag". That way if you change the passcode it is not necessary to re-encrypt all the user data but only a small portion of stored keys. The UID key is hard coded into the main SoC (system-on-chip) and is the part of the CPU hardware security engine. This UID key is not accessible to the running code, so it is impossible to brute-force the Passcode key without the matching SoC hardware being involved in the process.

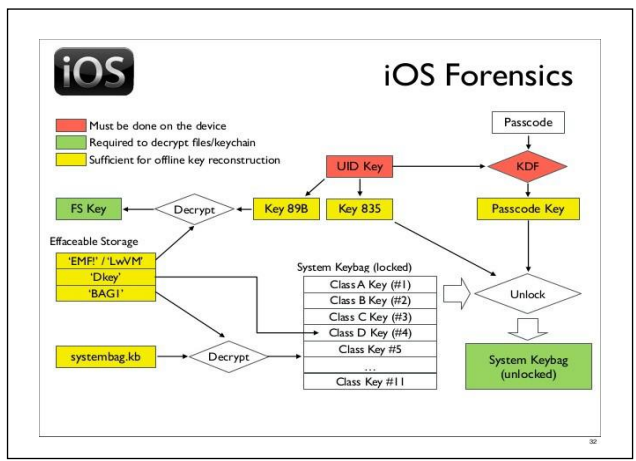

Figure 1. Block diagram of iOS encryption (Andrey Belenko [12])

If the security is enabled, then on powering up or waking up the iPhone asks for passcode (Figure 2). After 5 consecutive incorrect attempts a short waiting time of about 5 seconds is introduced; after the next incorrect attempt the time is increased to 1 minute; then 5 minutes; then 15 minutes and finally 60 minutes. There is an option to permanently delete all the data after 10 consecutive incorrect attempts.

Although information about iPhone 5c hardware is supposed to be secret, it can still be found on various related forums. This includes the layout of the main board as well as the circuit diagram or schematic and the bill of materials [13]. This significantly reduces the time otherwise required to locate important components such as A6 SoC and NAND. The pinout of the NAND chip is clearly outlined; however, some VCCQ and GND pins were swapped, thus simply following the circuit diagram could permanently damage the NAND chip. Flash memory devices in the LGA60 package [14] are not documented by any NAND chip manufacturers. Although present in their catalogs it was not possible to find any kind of documentation or datasheets on these particular NAND chips. This means that all the protocols and commands have to be learned by eavesdropping on the hardware interface using an oscilloscope or logic analyzer.

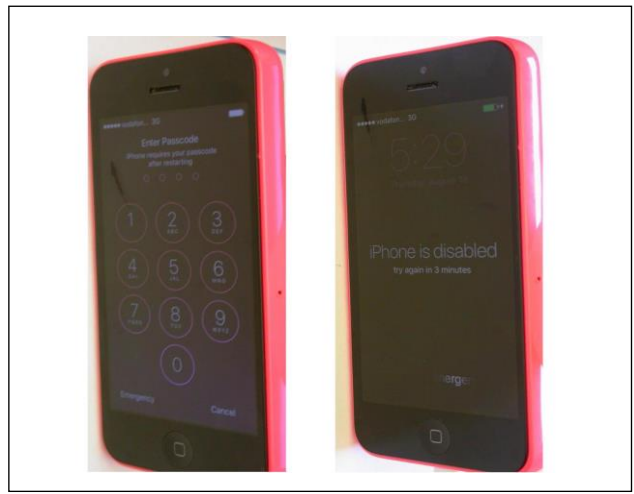

Figure 2. iPhone 5c passcode security

There are two major types of Flash memory – NOR and NAND [15]. In a NOR structure, memory cells are connected in a parallel manner, while in NAND the cells are connected in series which reduces the cell size. Also the highly regular NAND structure allows a much smaller fabrication process – 14nm versus 28nm for NOR. In addition, some NAND cells can store two or even three bits of information per cell. Such Flash devices are called MLC and TLC. The data transfer rate is also much higher than NOR. This makes this memory type the leader in high-density storage applications. However, NAND Flash memory has some drawbacks. First is the higher number of faulty cells which require external error correction. To help with that, NAND memory allocates additional space for error correction data. Second is the limited number of rewrites – usually tens of thousands versus hundreds of thousands for NOR. Also, NAND memory has a significant latency for accessing random memory blocks because the storage array is accessed sequentially.

Unlike magnetic media, which can be overwritten on the fly, Flash memory storage needs to be erased first before writing any new data. For example, SK hynix 64Gb (8G bytes) NAND Flash can be programmed in pages of 16384 data bytes plus 1280 error correction bytes [16]. However, it can only be erased in blocks of 256 pages or 4MB.

If some pages or blocks in NAND memory are no longer functional they have to be either replaced with fresh pages/blocks or marked as unusable. This can be achieved at the hardware level with a dedicated NAND memory controller, or done in software – either by the NAND memory driver or by the OS.

According to some sources, starting from iPhone 4 models, Apple started using a new type of NAND memory devices, called PPN (Perfect Page New) [17]. They use a dedicated memory controller that performs ECC (error code correction) on the fly thus reducing the load on the main SoC chip. This was achieved by acquiring the ECC controller manufacturer Anobit in 2011 [18]. Since then, four large NAND manufacturers Toshiba, SK hynix,

Samsung and SanDisk started production and supply of the new type of NAND memory devices in new LGA60 packages with custom interfaces.

Flash memory devices use different communication interfaces. They could be either serial or parallel. However, even serial SPI interface, could simultaneously run two or four data lines to increase the communication speed. Most NAND devices have a parallel interface with 8-bit or 16-bit width. Modern chips use DDR clocking to double the transfer rate, while some of them benefit from having more than one channel running in parallel. This not only increases the communication speed, but also allows erasing and writing in parallel. There is a standard agreed between all major NAND manufacturers called ONFi [19]. It outlines the electrical signals and specification as well as protocols and commands. As a first step in iPhone 5c NAND analysis it is necessary to check if the commands used for its NAND are the same as in the ONFi specification. There is another standard for parallel NAND Flash devices called eMMC [20]. It benefits from the embedded ECC controller and has a simpler interface compared to standard NAND.

### III. PREPARATION FOR MIRRORING

iPhone 5c devices are no longer manufactured, so samples for the hardware mirroring experiments were obtained on Ebay. Only two were fully functioning; the non-functioning ones were used as a test bench to verify ideas and finalize the best approaches. The fully functional phones were also updated to the latest 9.3 version of iOS.

#### A. TAKING IPHONE 5C APART

The teardown process of disassembling iPhone 5c is well described on the Internet with the iFixit web page probably being the best known [21]. After taking off the screen, unplugging all the connectors and carefully removing all the screws that hold the main board, it can be dismantled and taken off the frame. The image of the main board from both sides is shown in Figure 3.

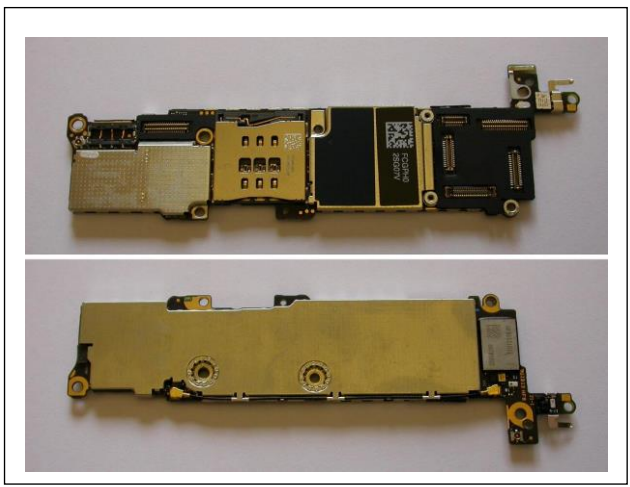

Figure 3. iPhone 5c main board

The next step is to gain access to the on-board components, in particular the NAND memory storage. Some shields need to be removed by desoldering them with a hot air gun. This exposes the A6 SoC chip on one side and the NAND chip on the other side (Figure 4).

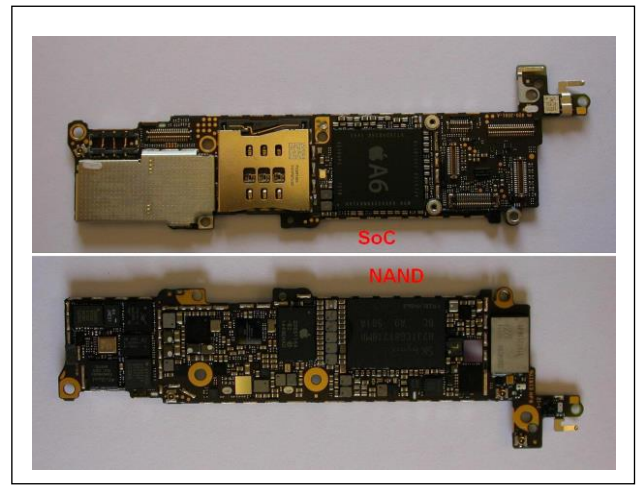

Figure 4. iPhone 5c opening for SoC and NAND

The NAND chip was not only soldered in its LGA leadless package but also glued using a strong epoxy compound. The gap between the NAND bottom side and the main PCB of merely 0.05mm makes the removal process quite a challenging task. Unless properly planned and supported by special thin blade knife tools this could damage both the NAND and the main PCB. To remove the NAND a temperature above 300 ºC is required due to heavy heat sinking of the main PCB. Although the surrounding epoxy becomes softer it still sticks to the nearby small components (capacitors, inductors etc.) and is likely to pull them off the main PCB, thus making it stop working. To avoid damage to other components, the epoxy was weakened along the perimeter of the NAND package with a Nichrome wire heated to about 700ºC by an electric current. Also small components around the NAND package were reinforced with high-temperature epoxy compound (Figure 5).

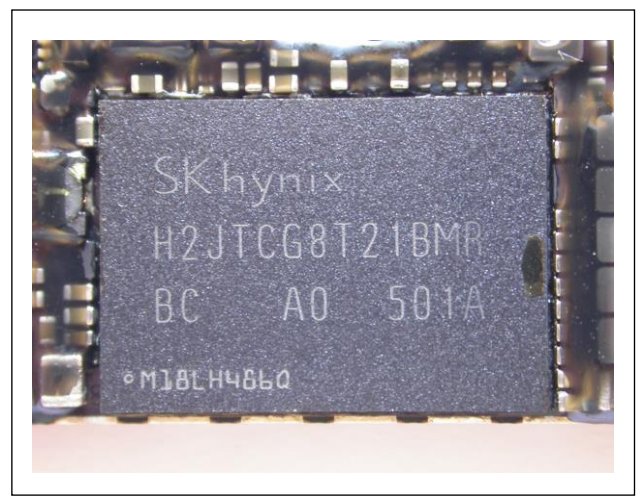

Figure 5. iPhone 5c NAND preparation

The process of removing the NAND chip started with fixing the main board in a holding frame. Then the whole board was preheated to about 150ºC before heating up the NAND package with 330ºC hot air. A thin-blade knife was used to slowly and carefully separate the NAND from the main board. The result is shown in Figure 6.

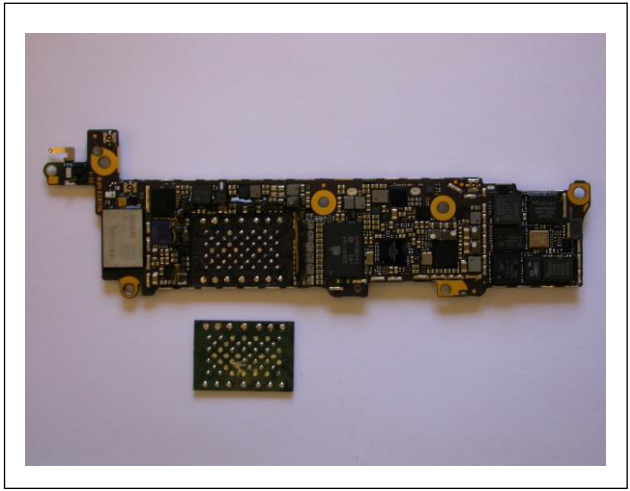

Figure 6.  iPhone 5c with removed NAND

Using high temperature air for desoldering the NAND chip should not cause any loss of data. Previous research demonstrated that Flash memory will sustain heating to temperatures as high as 400ºC for over ten minutes without any sign of degradation [22].

### B. WIRING NAND MEMORY

The next step was to check if the main board and NAND had survived the desoldering operation and still were fully functional. For that all the connected pads on the NAND package were connected to the corresponding pads on the main board using thin 0.3mm PTFE wires (Figure 7).

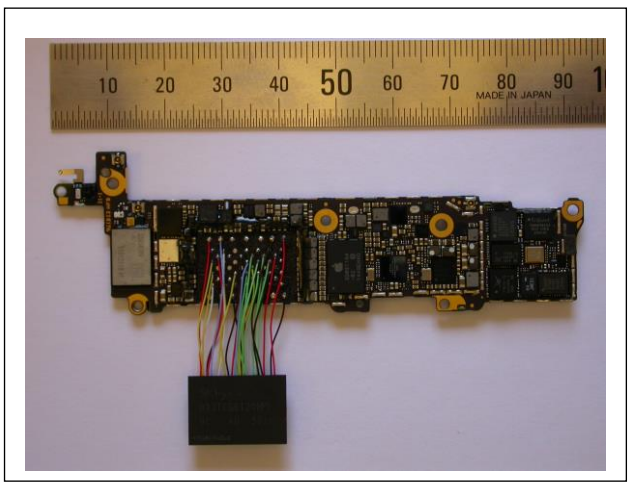

Figure 7.  iPhone 5c with wired up NAND

In order to test the result a hole was cut in the phone's frame as well as in the metal shield (Figure 8). Then both shields were soldered back to the main board before finally assembling the phone.

Once the phone was powered up the Apple logo initially appeared thus indicating that the NAND memory is functioning and the electrical connections are correct. Unfortunately, after 10 seconds it disappeared and then the boot process started over again. A quick look at the signals with an oscilloscope revealed that the power lines were not coping with the peak current during fast DDR communication. Therefore, several bypass capacitors were installed directly on all power pads of the NAND package. However, the NAND contents was already corrupted and the phone was still crashing after about 10 seconds into the boot process. Any attempts to force the iPhone into recovery mode, so that the NAND Flash can be reinstated, using iTunes were unsuccessful. The solution came with hard wiring both HRESET and FORCE_DFU signals off the SoC chip, by soldering thin wires to the corresponding pads on the main PCB and using switches to control those signals (Figure 9). HRESET signal is activated with a low logic level, but FORCE_DFU signal requires a higher voltage of 3V to be enforced. This was figured out by measuring the protection diodes between the power lines and the signals.

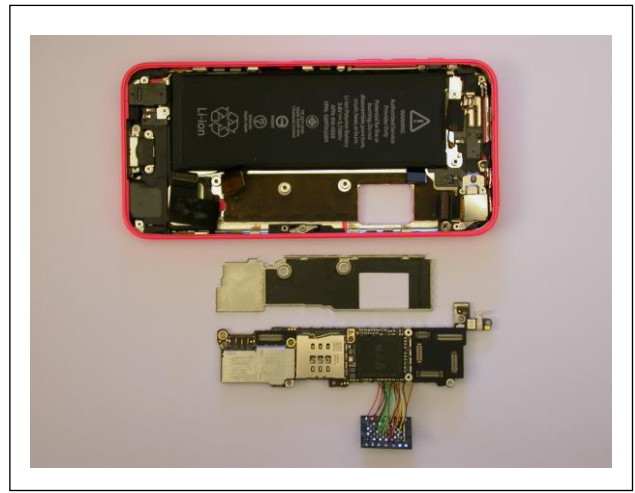

Figure 8.  iPhone 5c prepared for assembling

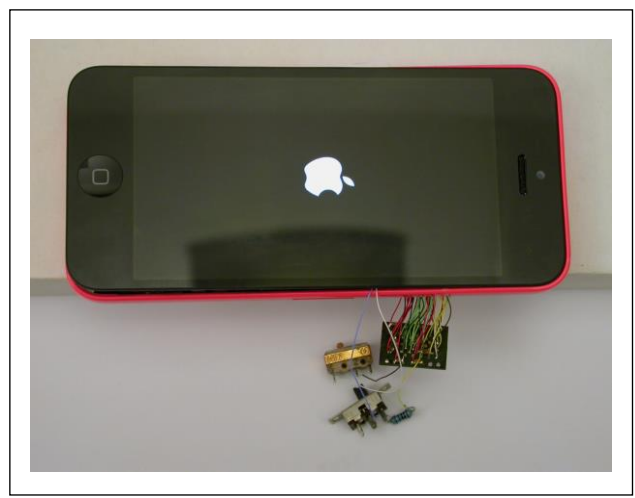

Figure 9.  iPhone 5c with hardware DFU forcing

After restoring the phone to its factory settings using iTunes and making sure that all functions work correctly, it was possible to continue the experiments. In order to assist with signal analysis, the NAND was wired to a small 0.05" connector, while the matching connector was soldered to a prototype PCB and wired to the main board of the iPhone 5c (Figure 10).

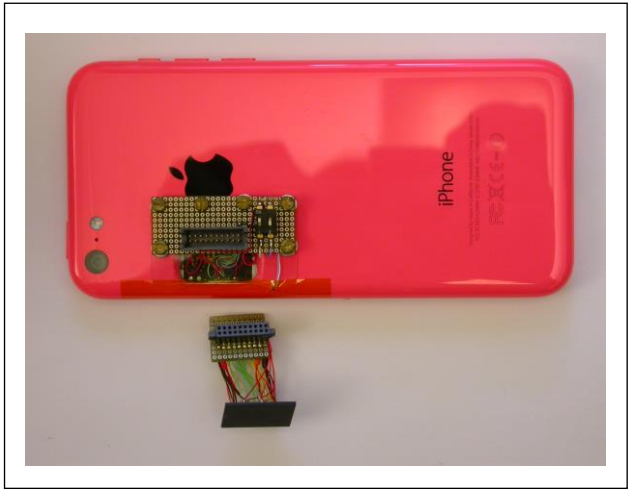

Figure 10. iPhone 5c with NAND on connector

Unfortunately, the phone did not work reliably with the NAND on a connector and was often crashing during the boot process, while any attempts to restore it with iTunes resulted with "An unknown error occurred (14)". Again an oscilloscope helped in finding that the communication signals were severely distorted because of the parasitic capacitance and inductance caused by the long wires (Figure 11). The top waveform is the source at the SoC side, while the bottom one shows the NAND side. The overshoot for 1.8V CMOS logic signal was +1.1V and the undershoot signal was −0.9V. This was a serious problem especially for clocking signals such as RE, WE and DQS, because the ringing was causing unwanted data latching.

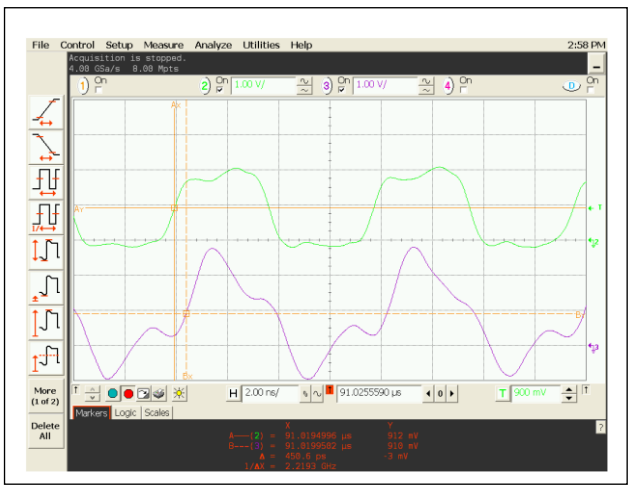

Figure 11. iPhone 5c signals at SoC and NAND pads

The problem was solved with insertion of small termination resistors into all signal lines. The resulting signal is presented in Figure 12. Only a small delay of 0.12ns was introduced as a result.

This again made the phone work perfectly and it was ready for the next step – of eavesdropping on the communication protocol and commands.

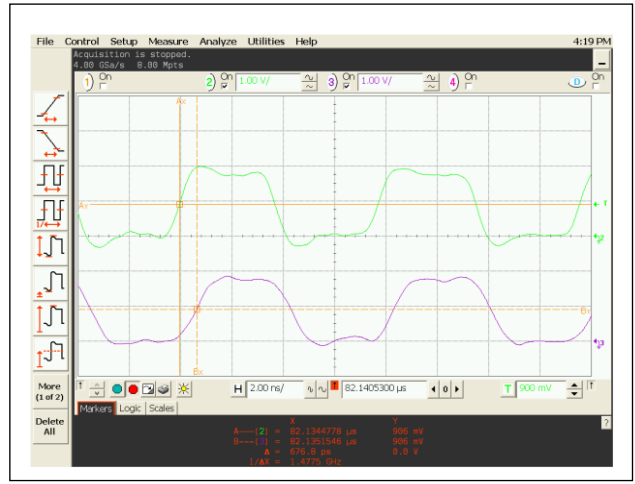

Figure 12. iPhone 5c signals after termination

### C. EAVESDROPPING ON NAND COMMUNICATION

In order to reliably eavesdrop on the communication an intermediate prototype PCB board was built with buffers (Figure 13). That way both an oscilloscope and logic analyzer probes can be used for signal acquisitions without fear of overloading the NAND communication signals with the high capacitance and low resistance of the probe leads.

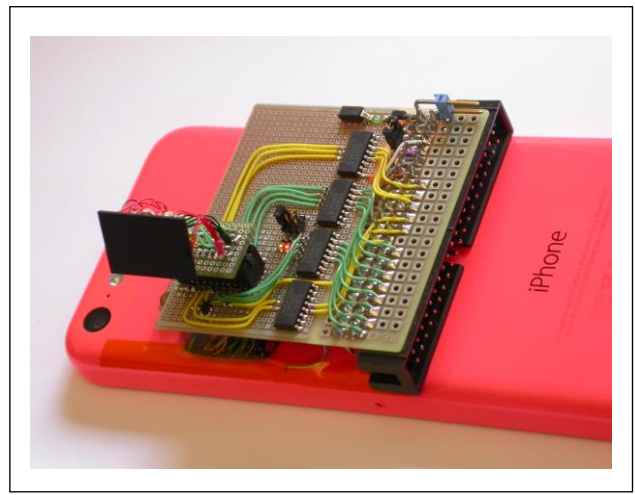

Figure 13. iPhone 5c with intermediate board for eavesdropping

Unfortunately, once again the signal integrity problem was encountered. This time it was caused by the input and output capacitance of the buffer elements. Standard digital CMOS logic of 74LVC and 74AVC series introduces a capacitance load in the region of 4pF to 6pF. This in combination with the inductance of the wires caused excessive delays and ringing in the signal lines. The only solution was to choose a logic with lower capacitance, such as the 74AUP and 74AXP series. However, by the time the capacitive load problem was solved, the NAND contents was corrupted beyond the iTunes capability to restore the iPhone with "An unknown error occurred (4013)" being reported all the time. The solution was again hardware-based. By carefully adjusting the delay on the RE line with an RC chain (Figure 14), it was possible to trick iTunes into thinking that the NAND storage is partially corrupted and force restoring both the LLB and iBoot partitions.

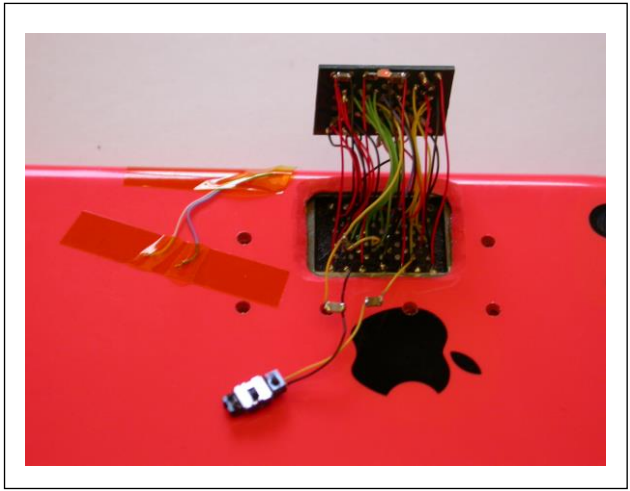

Figure 14. iPhone 5c with RC delay line for NAND

Finally the iPhone 5c was ready for examination of the protocols and commands during its boot process as well as during normal operation.

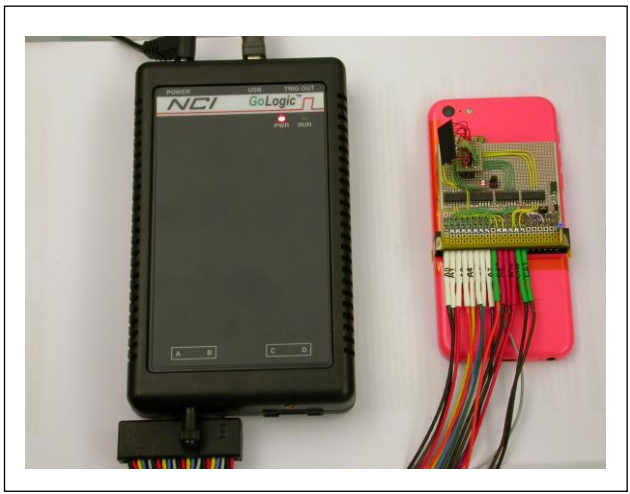

Figure 15. iPhone 5c with logic analyzer

The signal analysis with an oscilloscope revealed that the iPhone 5c uses different interfaces and commands at different stages of the boot process. At the very beginning it reads pages using ONFi compliant commands at a relatively low speed of 17MHz, then it switches to undocumented proprietary commands with 50MHz clocking. The iOS is loaded with a DDR clock of 128MHz thus achieving a peak throughput of above 250MB/s. The particular SK hynix H2JTCG8T21BMR 8GB NAND chip used in a sample phone has only one communication bus. NAND chips from other manufacturers use two buses and will require two sockets for mirroring.

A logic analyzer was used to record all the commands and work out the proprietary custom protocol used for the NAND communication in iPhone 5c (Figure 15).

## IV. IMPLEMENTATION OF MIRRORING

The next step was to implement all the necessary commands to support reading, erasing and writing of the Flash memory in a separate setup controlled by a PC via a serial port.

### A. PROTOCOL IMPLEMENTATION

In order to debug all the custom commands used for NAND communication a simple adapter cable was made for plugging the intermediate board with the NAND chip into a self-made universal IC programmer (Figure 16).

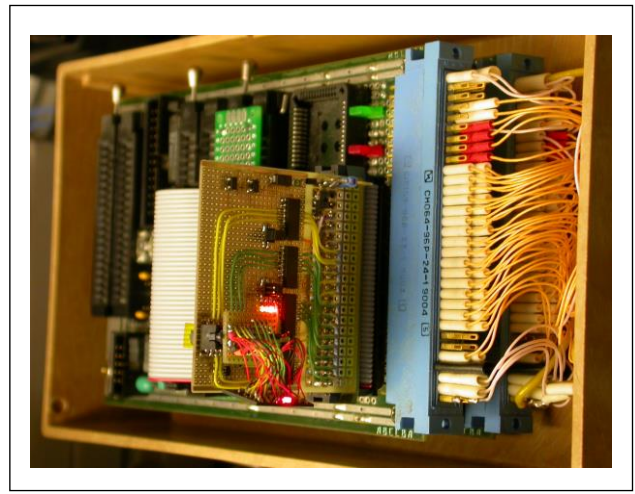

Figure 16. Evaluation of NAND in IC programmer

Figure 17. iPhone 5c NAND reading software

All the signals were replicated with a slower communication speed of 1MHz using the C programming language. This helped a lot in understanding the layout of the memory. For 8GB storage there are two planes each containing 1064 blocks. Erase operation can only be applied to blocks. Each block contains 256 pages and the writing is done in pages. Each page contains 16448 bytes of information. This information is grouped in four sectors of 4096 bytes of data and 16 bytes of indexing. It can be seen in Figure 17 that the ASCII text is interrupted with 16 bytes of binary data from 0x3430 to 0x343F. Very likely the indexing is used to mark the logical mapping of the data for wear levelling of the Flash memory. This is because the physical addresses in NAND memory are constantly changing to avoid premature damage of Flash cells which can only be rewritten a limited number of times.

### B. BACKING UP

Due to the high capacity of the NAND storage it was most convenient to use the same type of memory to hold the backup copy. For that the same SK hynix 8GB type of the NAND chip was desoldered from a non-working iPhone 5c board and wired to a connector. Obviously, because of the different UID in that iPhone, the chip did not operate in the test phone. Nevertheless, iTunes did allow us to restore the system to the initial state although with a different serial number. However, the iPhone failed the activation stage probably because of the changed serial number. Nevertheless, this NAND chip was still useful as backup storage.

In order to create the exact backup copy of the NAND chip a special test board was built (Figure 18). The core of the board is the Microchip PIC24EP512GP806 microcontroller that has a hardware PMP port capable of running at 40MB/s transfer speed for reading and 80MB/s for writing [23]. At such throughput it takes about 80 minutes to copy all the data from the original chip to the backup. Unfortunately, the 1:1 backup copy did not work in the iPhone 5c. Even the boot Apple logo did not appear on the power up. There were some references to hidden partitions used in iPhone NAND storage which makes cloning a challenging task [24].

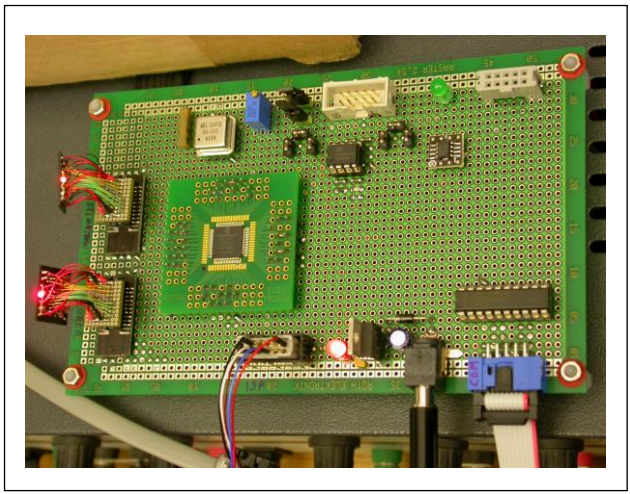

Figure 18. Test board for copying NAND chips

The backup copy was then used to restore the data in the original NAND chip after several passcode attempts.

### C. RESTORING

The process of NAND mirroring is relatively simple. Once the backup copy is created and verified, the original chip is plugged back into the iPhone 5c. After the power up, which takes about 35 seconds, we enter the passcode 6 times. Then the phone is powered down by holding the power button and sliding the power off message. It is necessary to wait until the power is removed from the NAND which takes about 10 seconds. Once the LED attached to the NAND goes off it is safe to remove the NAND and plug it into the test board.

Our software that was running on a laptop first scanned the areas of the most likely changes in the NAND and created a file with checksums. This file was then compared against the backup scan. All the changed blocks were then erased and all used pages were written back from the backup copy (Figure 19). This process takes from 30 to 60 seconds depending on memory usage. Once the restore is completed the test board can be powered off and the NAND can be placed back into the phone.

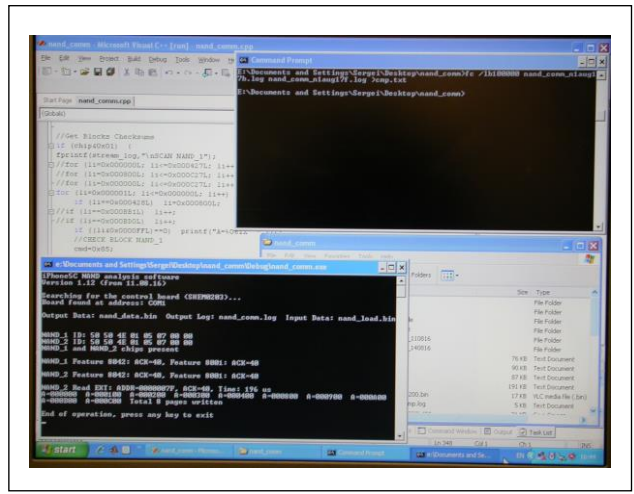

Figure 19. iPhone 5c NAND mirroring software

Once the phone is powered up and the screen is slid the passcode can be entered six times until the delay of one minute is introduced again. Then the process of mirroring from backup can be repeated again and again until the correct passcode is found. On average each cycle of mirroring for six passcode attempts takes 90 seconds. Hence, a full scan of all possible 4-digit passcodes will take about 40 hours or less than two days.

It is not easy to describe the whole process in a paper, therefore a video of the working proof-of-concept demonstration for this NAND mirroring process will be placed on the Internet. More information can be found on the dedicated iPhone research project page [25].

In the presented method the original chip is always restored to the initial passcode attempts counter state without applying wear levelling. As a result, its Flash memory gets worn out. Although NAND chips allow a few thousand rewrites, no one knows how many were used already. Hence, it could fail before the correct passcode is found. Given six attempts per each rewrite this method would require at most 1667 rewrites to find a 4-digit passcode. For a 6-digit passcode it would require over 160 thousand rewrites and will very likely damage the Flash memory storage.

From a forensics point of view modifying the original NAND storage will be undesirable because this could change some vital information in the device.

### D. CLONING

The process of cloning involves creating a fully working copy of the NAND Flash memory chip. However, as it was already mentioned in the previous section, simply copying the 8GB information from the original chip into another identical chip taken from other iPhone 5c does not give the desired result and the iPhone does not boot.

```
00041A00:61:0045        00841A00:40:0045
00041A01:61:0AB3        00841A01:40:0AB3
00041A02:49:FFFF        00841A02:40:ED0F
00041A03:61:ED0F        00841A03:40:F049
00041A04:49:FFFF        00841A04:40:02AA
00041A05:61:F049        00841A05:40:E4B5
00041A06:49:FFFF        00841A06:40:2A2C
00041A07:61:02AA        00841A07:40:BFBA
00041A08:49:FFFF        00841A08:40:E2B4
00041A09:61:E4B5        00841A09:40:01B9
00041A0A:49:FFFF        00841A0A:40:D5B0
00041A0B:61:2A2C        00841A0B:40:FCAB
00041A0C:49:FFFF        00841A0C:40:E14E
00041A0D:61:BFBA        00841A0D:40:E6FE
00041A0E:49:FFFF        00841A0E:40:E3EA
00041A0F:61:E2B4        00841A0F:40:DF68
00041A10:49:FFFF        00841A10:40:E305
00041A11:61:01B9        00841A11:40:F389
00041A12:49:FFFF        00841A12:40:D386
00041A13:61:D5B0        00841A13:40:0AB2
```

Figure 20. iPhone 5c hidden pages in NAND memory space

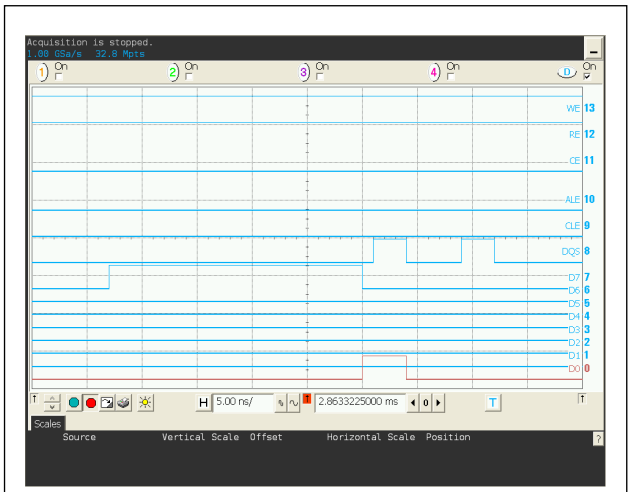

Figure 21. iPhone 5c waveforms during NAND access

Some additional research was undertaken to figure out why simple copying does not work. For that the same model of the NAND Flash chip was programmed with the data from the original chip and then the communication was analyzed with both an oscilloscope and a logic analyzer. First, some pages were accessed from addresses outside the normal 16GB space. For example, instead of reading and writing to the block 0x00041Axx the CPU was accessing the block 0x00841Axx. Although such accesses are mapped back into the 0x00041Axx block, the page numbers were different as well as the status of those pages. Figure 20 shows status and checksums for the first several pages for the blocks 0x00041Axx and 0x00841Axx. It can be noted that the page 0x00841A02 is mapped to 0x00041A03, page 0x00841A03 to 0x00041A05 and so on. The status of the pages for 0x00041Axx block was 0x61 in comparison with 0x40 for 0x00841Axx block. Second, some irregularities were found in the communication prior the access to those hidden pages (Figure 21). Although the data transfer during the access is performed in the SDR mode at 17MHz, while the configuration commands use even the slower speed of about 1MHz, some data inside the commands are smuggled at an astonishing rate of 256MB/s in DDR3 mode. Also, a dummy value for data bit 7 was introduced for the period of 23ns. Given that the data setup time in those transfers is less than 1ns there is a very high chance that such information would be overlooked by most would-be attackers.

After those findings the implementation of the communication protocol was amended in the test board. Then the data-mirroring software was modified to include cloning the hidden pages. As a result the newly created clone of the original NAND chip was fully functional in the iPhone 5c. It was then tested with six incorrect passcode attempts before replacing it with the original chip. After the boot process it was possible to enter the incorrect passcodes again six times until the one minute delay was introduced. This fully proved the correctness of the hardware NAND mirroring attack on iPhone 5c.

Because there is no limitation on the number of such NAND clones, they can be created in advance and restored in parallel when one of them is being used for passcode testing. This way it only requires 45 seconds per six passcode attempts. For 4-digit passcode the maximum attack time would be $(10000/6) \times 45 = 75000$ seconds or about 20 hours. For 6-digit passcode this time will increase to about 3 months which in some cases might be acceptable.

## V. IMPLICATIONS

There are several limitations to the presented methods. For attacking real devices some form of automation for the process is necessary to avoid mechanically plugging of the NAND memory devices into the iPhone. This could be achieved either using electrical switches or multiplexers, or by emulating the NAND chip, for example, using an FPGA.

Another concern is that the process of removing the NAND chip from the main board could damage both the main board and the NAND Flash chip.

Some reliability issues were found in the NAND management. When NAND chips from non-working iPhone 5c main boards were desoldered and tested, some of the chips had bad blocks. Surprisingly, those bad blocks were primarily in two address spaces: 0x000BB100 – 0x000BB1FF and 0x000BD000 – 0x000BD0FF. The same addresses were overwritten with different data, while for other addresses the overwriting was happening into new addresses. The same applies to address spaces 0x0003B1xx and 0x0003D0xx. This means that the Flash wear levelling algorithm was not implemented correctly and some addresses are more likely to fail.

## VI. FUTURE WORK

The iPhone 5c device being analyzed in this research project was far from the latest Apple phones. Since then several new models were introduced such as iPhone 5s, iPhone 6 and 6s, iPhone SE and iPhone 7. However, iPhone 5s and 6 use the same type of NAND Flash memory devices. It would be logical to test them against mirroring. For models from iPhone 6s more sophisticated hardware will be required because they use high speed serial NAND Flash chips with a PCIe interface.

Some improvements to the existing setup could involve automating passcode entry and rebooting, for example, by using an external USB controller to emulate all the necessary functions for passcode entry.

Having taken this technology to proof-of-concept, it would be useful to develop it further by building a fully working emulator of the NAND Flash memory chip. This would significantly speed up the mirroring process. Further improvement could come from finding a way for soft restart or placing the iPhone into sleep mode. This would save the time otherwise needed for a full boot of the iOS.

It would be beneficial to develop a safer way of removing the NAND Flash chip from the main board, or a way of reading out the NAND Flash contents without the need to physically remove it. The NAND chip has two additional pads which might be used to gain access to the internal memory contents. Those pads are wired to test pads on the main board. This could be a backdoor to the NAND data. Of course, there will be a challenge to figure out the exact interface and the debugging protocol. One possible approach could be in employing advance hardware silicon scanning technique presented a few years ago [26].

The research presented in this paper had a limited budget and the time constraints. With a properly funded research project and a decent research team the evaluation of the NAND storage in mobile devices could be taken to a new level.

VII. CONCLUSION

The research presented in this paper has demonstrated the successful hardware mirroring of NAND Flash memory in the Apple iPhone 5c. It was possible to bypass the passcode retry limit by restoring the original NAND data from a previously created backup copy. Then the mirroring process was improved by creating a fully working clone of the original chip. That way the forensics of the original chip are fully preserved and multiple copies of the original can be used for brute-forcing the passcode. This is the first public demonstration of a working prototype and real hardware mirroring for the iPhone 5c. It should allow brute-forcing of a 4-digit passcode in less than a day.

There were several obstacles on the way. Although Apple did introduce some hardware security countermeasures, most of them seem more like security-through-obscurity rather than fully thought through measures.

The experimental setup presented in this paper could also be used as a test platform to evaluate and observe NAND memory in real time. This might help in spotting not only the hardware security related issues but also some reliability issues. For example, it was found that half of the NAND chips from non-working iPhone 5c main boards had specific blocks failed due to excessive rewriting. This might happen because of a bug in Flash memory wear levelling algorithm as it was implemented in software.

There are several ways the presented approach could be improved. First, the NAND access can be optimized using more efficient hardware and software support. Second, the whole process of passcode recovery can be automated. Third, instead of using the original chip a proper 1:1 clone can be created. Finally, the NAND memory storage can be emulated in an FPGA thus eliminating the need to waste time rewriting the NAND Flash. The last two improvements will be particularly useful for forensic applications where the original NAND storage must be preserved. Ultimately, the NAND might be read without taking it off the main board. That way the risk of damaging the data will be minimal.

Further research should be undertaken to understand the risks involved in NAND storage mirroring in other mobile devices.

The mirroring solution presented in this paper was achieved using off-the-shelf components bought from an electronics distributor with a budget of under 100 dollars. The same approach could be applied to the newer models of iPhone. The same type of LGA60 NAND chips are used up to the iPhone 6 Plus. Any attacker with sufficient technical skills could repeat the experiments. Newer iPhones will require more sophisticated equipment and FPGA test boards.

In terms of countermeasures against mirroring, several approaches can be taken at various levels. At the hardware level more robust authentication should be used rather than a proprietary interface. At the software level, a challenge-response authentication could be used to prevent access to NAND memory or replay attacks. At the usability level, users should use at least 6-digit passcodes, or better 8-character passcodes. Attacking such passcodes would require access to the SoC directly to reduce the waiting time between attempts.

The knowledge of the possibility of mirroring would definitely help in designing systems with better protection.

Despite government comments about feasibility of the NAND mirroring for iPhone 5c it was now proved to be fully working.

ACKNOWLEDGEMENT

I would like to thank Dr Markus Kuhn for providing the working iPhone 5c sample used in my experiments and for his helpful discussions, and to Prof Ross Anderson for testing the attack at our Security Group meeting.